\documentclass{iau} 
\usepackage{graphicx}

\title[JD 11.~Abundances and ages of the outer regions of spiral disks] 
{Properties of the outer regions of spiral disks: abundances, colors and ages}

\author[Moll{\' a} et al.]   
{Mercedes Moll{\'a}$^1$, Angeles I. D{\'\i}az$^{2,3}$, Brad  K. Gibson$^{4}$, Oscar Cavichia$^{5}$,
and {\'A}ngel-R. L{\'o}pez-S{\'a}nchez$^{6,7}$}

\affiliation{$^1$Departamento de Investigaci\'{o}n B\'{a}sica, CIEMAT, 28040, Madrid. Spain
\\email: {\tt mercedes.molla@ciemat.es} \\[\affilskip]
$^{2}$ Universidad Aut\'{o}noma de Madrid, 28049, Madrid, Spain \\
$^{3}$ Astro-UAM, Unidad Asociada CSIC, Universidad Aut{\'o}noma de Madrid, 28049, Madrid, Spain\\
$^{4}$ E.A. Milne Centre for Astrophysics, Dept. of Physics \& Mathematics, University of Hull, Hull, HU6~7RX, United Kingdom\\
$^{5}$ Instituto de F\'{i}sica e Qu\'{i}mica, Universidade Federal de Itajub\'{a}, Av. BPS, 1303, 37500-903, Itajub\'{a}-MG, Brazil\\
$^{6}$Australian Astronomical Observatory, PO Box 915, North Ryde, NSW 1670, Australia\\ 
$^{7}$Department of Physics and Astronomy, Macquarie University, NSW 2109, Australia}

\pubyear{2015}
\volume{321}  
\jname{Formation and evolution of galaxy outskirts}
\editors{Armando Gil de Paz, Johan Knapen \& Janice Lee }
\begin{document}

\maketitle

\begin{abstract}
We summarize the results obtained from our suite of chemical evolution 
models for spiral disks, computed for different total masses and star 
formation efficiencies.  Once the gas, stars and star formation 
radial distributions are reproduced, we analyze the
Oxygen abundances radial profiles for gas and stars, in addition to stellar averaged ages
and global metallicity. We examine scenarios for the potential origin of
the apparent flattening of abundance gradients in the outskirts of disk 
galaxies, in particular the role of molecular gas formation 
prescriptions.
\end{abstract}
\firstsection
\section{Introduction}

Chemical evolution models are the classical tool by which to interpret 
observed elemental abundances, and associated quantities such as gas and 
stellar surface densities, star formation histories, and the 
distribution of stellar ages.  Elemental patterns carry the fingerprint 
of star formation timescales from their birth location, regardless of a 
star's present-day position.  Chemical evolution codes solve a system of 
first order integro-differential equations, assuming an analytical star 
formation (SF) law, initial mass function (IMF), stellar lifetimes, and 
nucleosynthetic yields.
 
In \cite{md05}, we calculated a grid of 440 theoretical galaxy models, 
(44 radial mass distributions, and 10 molecular gas and SF efficiencies 
between 0 and 1), calibrated on the Milky Way Galaxy (MWG). SF was 
assumed to occur in two steps: 1) molecular clouds forming from diffuse 
gas; 2) cloud-cloud collisions creating stars. Radial distributions for 
{\bf both} gas phases were derived, but the inferred predicted ratios of 
atomic to molecular gas, and SF rate ($SFR$), were found to be at variance 
with those observed. We are computing a new grid of models with updated 
stellar yields --\cite{molla15}--,  gas infall rates --\cite{molla16a}--, and  
molecular gas formation efficiency --\cite{molla16b}.
Our aim is to improve the predicted $H_{2}$ and $SFR$ profiles, while 
maintaining abundance radial gradients in agreement with those observed. 
We summarize our updated models and results in \S2 and our
conclusions in \S3.

\section{New chemical evolution models}

As described in \cite{molla16a}, we compute the radial mass distributions 
for 16 theoretical galaxies, following \cite{sal07}, who define them in 
terms of $M_{vir}$ and their associated rotation curves.  The virial 
masses, defined as the total dynamical mass for a galaxy, are in the range $M_{vir} \in [5\times 10^{10} - 
10^{13}]\,M_{\odot}$, with associated disk masses in the range $M_{disk} 
\in [1.25\times 10^{8} -5.3\times 10^{11}]\,M_{\odot}$. The initial gas 
in each model collapses onto the disk on timescales based in 
\cite{sha06}, which gives the ratio between the disk and the virial 
masses, $M_{disk}/M_{vir}$. From the rotation curves, we calculate the 
radial distribution of the dynamical mass and the one that the disk will 
have at the present time. Thus, we obtain the infall rate necessary to 
have, at the end of a model's evolution, the appropriate disk for each 
dynamical mass. The inferred infall rates evolve modestly with time for 
the disks (stronger for bulges), showing, among radial regions or among galaxies, 
only variations in the absolute values. The radial regions in a disk for a galaxy with 
$M_{vir}\sim10^{12}$~$M$$_\odot$ (i.e., a MWG-like analog), have infall 
rates at the present time $\dot{M}\sim 0.5\,M_{\odot}\,yr^{-1}$ 
for galactocentric radii $R<13$\,kpc, in agreement with \cite{san08}'s 
data, while it is much lower in the outer regions ($R> 13$\,kpc). 

By following \cite{molla15}, we use the stellar yield sets from 
\cite{lim03,chi04} for massive stars combined with the IMF from 
\cite{kro}, joined to yields from \cite{gav05,gav06} for the low and 
intermediate mass stars.

\begin{figure}
\begin{center}
\includegraphics[width=5cm, angle=-90]{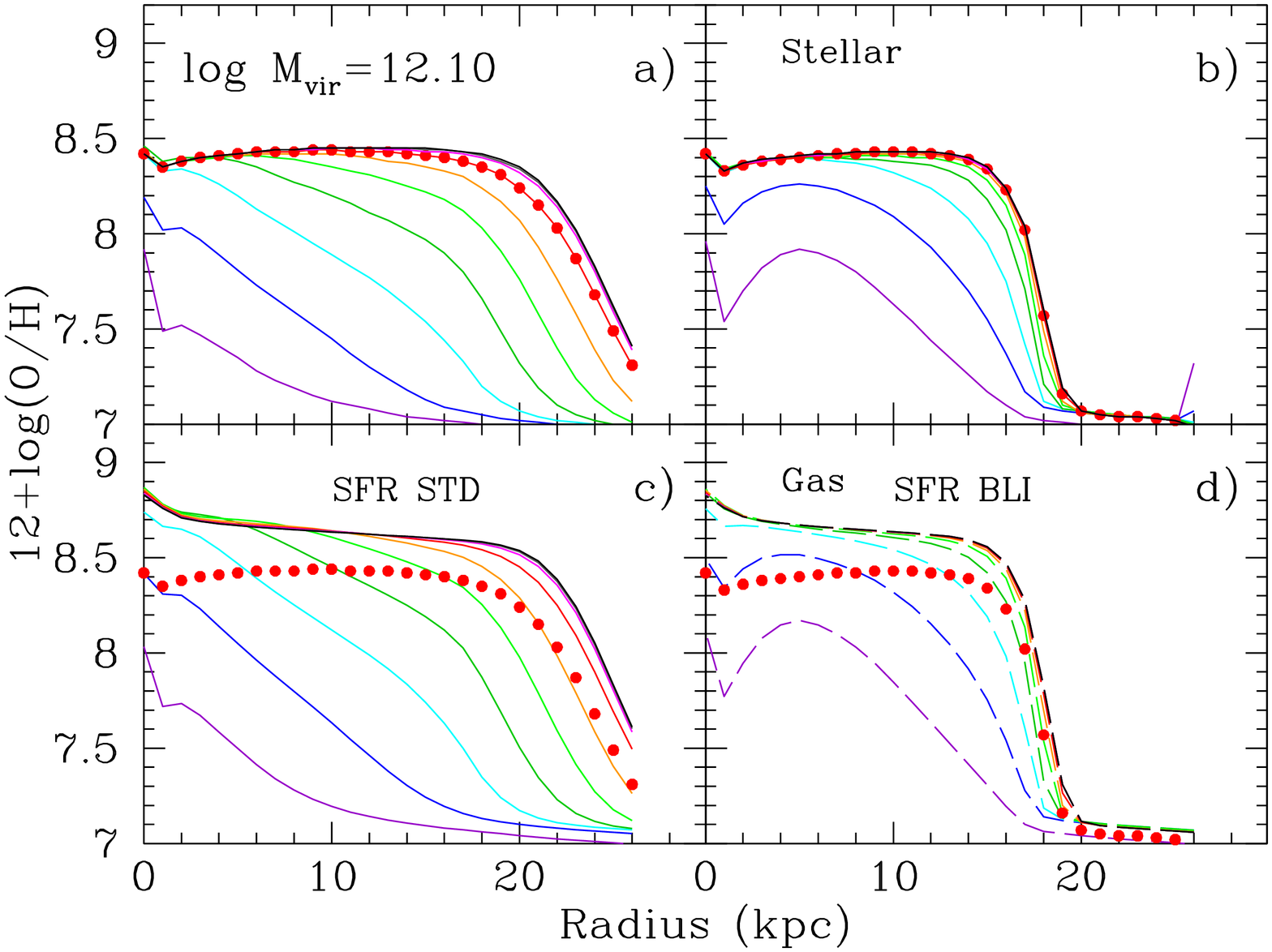} 
\includegraphics[width=5cm, angle=-90]{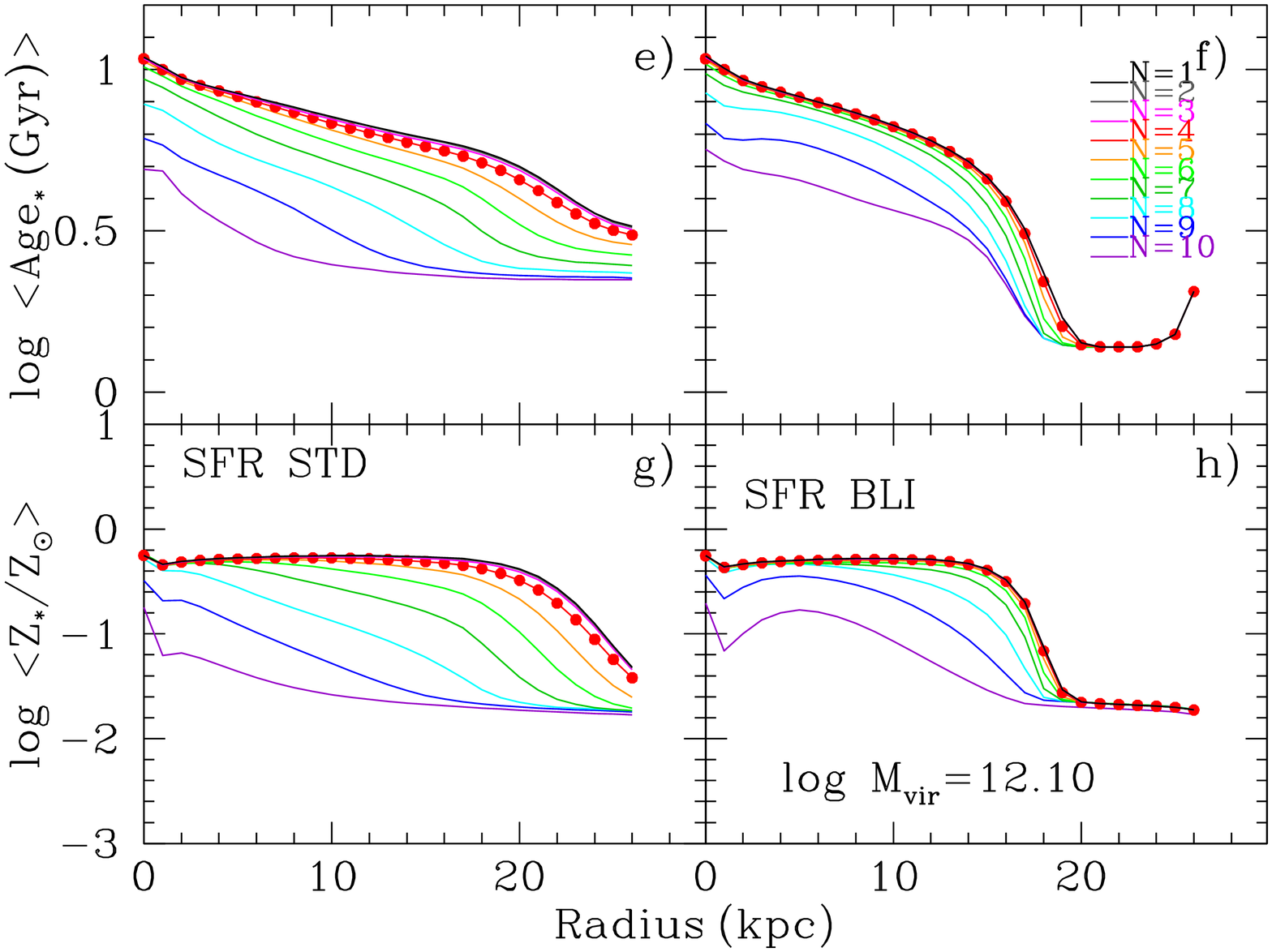} 
\caption{Radial distributions of oxygen abundances for
stars (panels a and b) and gas (panels c and d) obtained with STD (panels a and c) and BLI models (panels b and d). 
Radial distributions of stellar average age, in logarithmic scale  in e) and f); 
and averaged metallicity, $<Z/Z_{\odot}>$, in g) and h), 
obtained with STD, e) and g) and BLI, 
f) and h), models. Each color shows a different SF efficiency value. See text for more explanations.}
\label{age-Z}
\end{center}
\end{figure}

\begin{figure}
\begin{center}
\includegraphics[width=5cm, angle=-90]{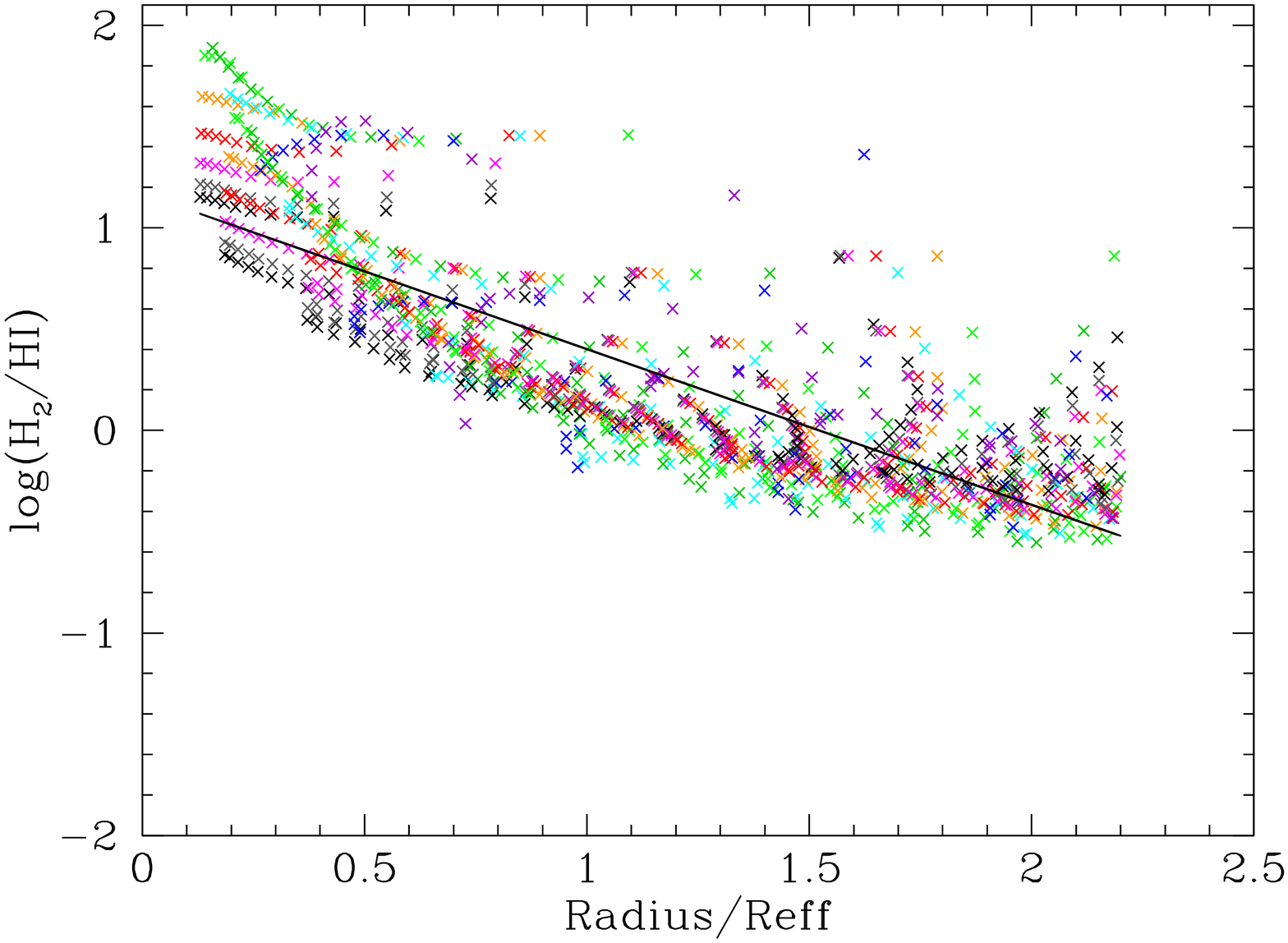} 
\includegraphics[width=5cm, angle=-90]{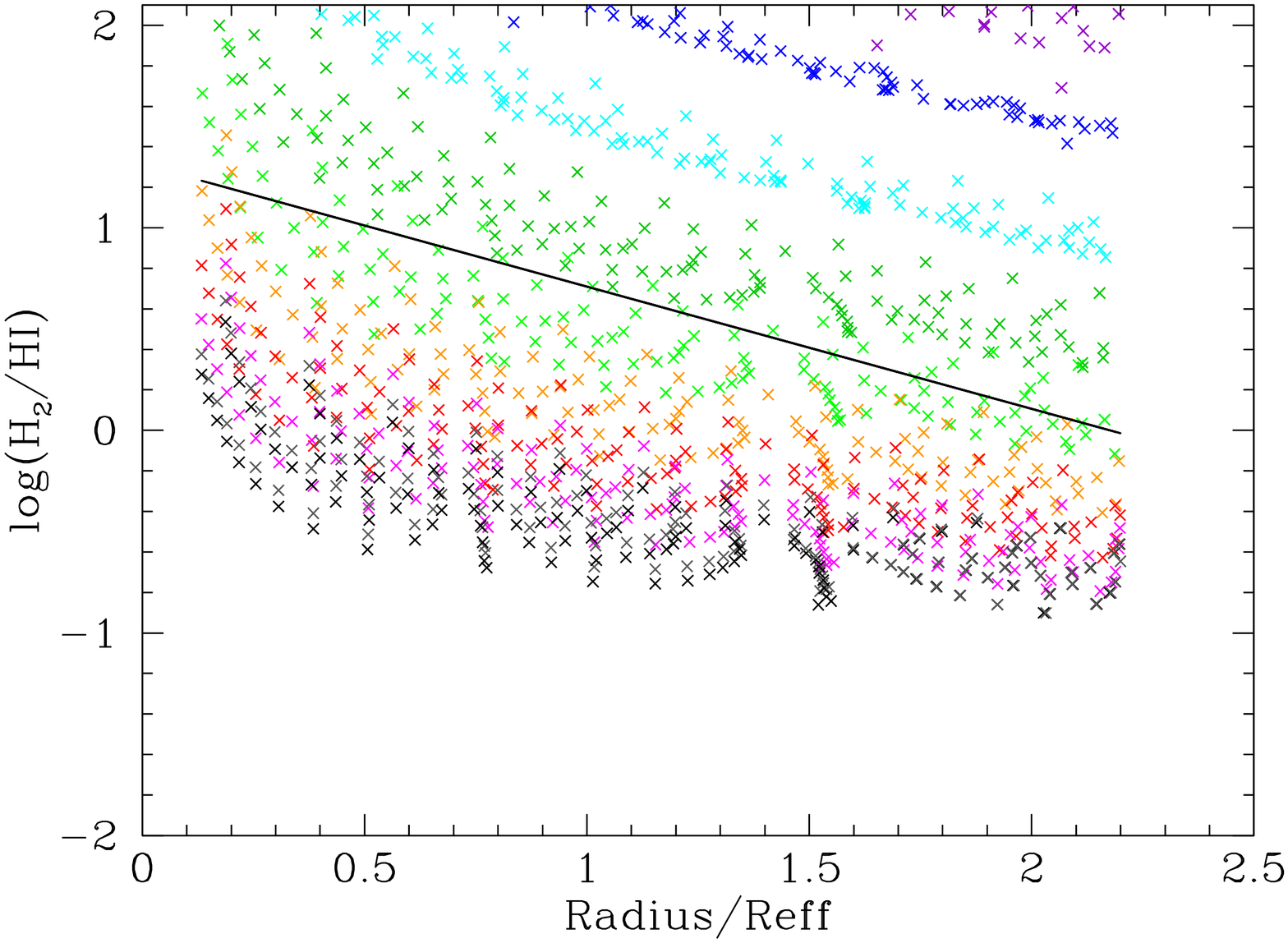} 
\caption{The ratio $HI/H_{2}$ {\sl} versus the normalized radius 
$R/R_{eff}$, for the 76 galaxy models and 10 values of efficiencies to 
form stars in the: left) STD; right) BLI models.}
\label{profiles}
\end{center}
\end{figure}

Molecular gas in our models is created from diffuse gas. The efficiency 
of this process takes a value between 0 and 1, as a probability factor, 
in our standard models (STD), with an exponential function: 
$\epsilon_{c}= exp(N^{2}/20)$. With $N=4$ we obtain a MWG-like model 
that produces a good fit for the evolution of the solar region, and for 
the radial distributions of gas, stars, SFR, and elemental abundances of 
C, N, and O, as shown in \cite{molla15}. We now check, in \cite{molla16b}, the prescriptions 
in creating molecular gas from \cite{fu10}, based upon \cite{bli}: the 
$H_{2}$ fraction depends on total pressure, which, in turn, depends on 
gas and stellar surface densities. This model (BLI) is contrasted with 
STD.  The dependence on stellar density produces a threshold effect, and 
in this way the evolution is slow at the beginning, but stronger at 
later times in the BLI model, relative to the STD.  This produces a 
steepening of the radial distributions: the gradients of oxygen in stars 
and gas, $12+log(O/H)$, stellar age $\log {<Age>}$\,(Gyr), and 
metallicity $<Z/Z_{\odot}>$ are steeper in the BLI panels (b,d,f,h), with 
a strong flattening in the outer disk, compared with the STD results 
(even showing a U-shape in age in the outermost regions, in the absence 
of any stellar radial migration) at the last radial region). Finally, we 
show in Fig.~\ref{profiles}, the normalized radial distribution of the 
molecular gas fraction obtained for our new models, using the STD and 
BLI options, compared with the empirical relationship obtained from 
\cite{bigiel08}. The BLI models, while showing reasonable global trends, 
do not fit the data well.

\section{Conclusions}
\begin{itemize}
\item A grid of chemical evolution models with 16 dynamical masses in 
the range 10$^{10}$ to 10$^{13}\,M_{\odot}$ is calculated. A MWG-like 
model reproduces very well the observed radial distributions, as shown 
in \cite{molla15} and \cite{molla16b}.
\item Prescriptions from \cite{bli} and \cite{fu10} for the formation of H$_{2}$ 
(BLI models) produce radial variations in O, $<Age>$, and $<Z/Z_{\odot}>$ which are stronger 
than STD models. Radial gradients are shown to be not invariant with 
radius.
\item The H$_{2}$/HI relationship from \cite{bigiel08} is obtained for 
STD models, while BLI shows an unrealistically high dispersion.
\end{itemize}

\end{document}